%% file: paper.tex
\documentclass[conference]{IEEEtran}
\usepackage{amsmath}
\usepackage{amsthm}
\usepackage{amsfonts}
\usepackage{amssymb}
\usepackage{graphicx}
\usepackage[keeplastbox]{flushend}
\usepackage{url}
\usepackage[T2A,T1]{fontenc}
\usepackage[font={footnotesize}]{caption}
\usepackage{color}    
\usepackage{threeparttable}
\usepackage{booktabs} 
\usepackage{multirow}
\usepackage{tabularx} 
\usepackage{msc}
\newcolumntype{L}[1]{>{\raggedright\arraybackslash}p{#1}}
\newcolumntype{C}[1]{>{\centering\arraybackslash}p{#1}}
\newcolumntype{R}[1]{>{\raggedleft\arraybackslash}p{#1}}
\usepackage{pgfplots}
\usepackage{scalefnt}
\usepackage[ruled,vlined]{algorithm2e}
\usepackage{paralist}
\usepackage{tikz}
\pgfdeclarelayer{bg}
\pgfsetlayers{bg,main}
\usetikzlibrary{decorations.text} 
\usetikzlibrary{backgrounds}
\usepackage[]{hyperref}
\usetikzlibrary{decorations.pathreplacing}
\usepackage{listings}

\usepackage{pifont}
\newcommand{\cmark}{\ding{51}}%
\newcommand{\xmark}{\ding{55}}%

\newtheorem{example}{Example}

\renewcommand{\tablename}{Tab.}

\renewcommand{\figurename}{Fig.}
 
\makeatletter\def\@captype{table}\makeatother

\input{code/solidity-highlighting.tex}
\colorlet{punct}{red!60!black}
\definecolor{delim}{RGB}{20,105,176}
\colorlet{numb}{magenta!60!black}
\lstdefinelanguage{json}{
    numbers=left,
    numberstyle=\scriptsize,
    stepnumber=1,
    numbersep=8pt,
    showstringspaces=false,
    breaklines=true,
    literate=
     *{0}{{{\color{black}0}}}{1}
      {1}{{{\color{black}1}}}{1}
      {2}{{{\color{black}2}}}{1}
      {3}{{{\color{black}3}}}{1}
      {4}{{{\color{black}4}}}{1}
      {5}{{{\color{black}5}}}{1}
      {6}{{{\color{black}6}}}{1}
      {7}{{{\color{black}7}}}{1}
      {8}{{{\color{black}8}}}{1}
      {9}{{{\color{black}9}}}{1}
      {:}{{{\color{punct}{:}}}}{1}
      {,}{{{\color{punct}{,}}}}{1}
      {\{}{{{\color{delim}{\{}}}}{1}
      {\}}{{{\color{delim}{\}}}}}{1}
      {[}{{{\color{delim}{[}}}}{1}
      {]}{{{\color{delim}{]}}}}{1},
}

\begin{document}

\newcommand{\smacs}{SMACS\xspace}%
\newcommand{\runtime}{runtime\xspace}%
\newcommand{\ts}{Token Service\xspace}%
\newcommand{\acr}{ACRs\xspace}%

\title{\smacs: Smart Contract Access Control Service
}

\author{
    \IEEEauthorblockN{Bowen Liu\IEEEauthorrefmark{1}, Siwei Sun\IEEEauthorrefmark{2},
    Pawel Szalachowski\IEEEauthorrefmark{1}}
    \IEEEauthorblockA{\IEEEauthorrefmark{1}Singapore University of Technology and Design, Singapore}
    \IEEEauthorblockA{\IEEEauthorrefmark{2}State Key Laboratory of Information Security, Institute of Information Engineering,\\Chinese Academy of Sciences, Beijing 100093, China}
}

\maketitle

\begin{abstract}
    Although blockchain-based smart contracts promise
    a ``trustless'' way of enforcing agreements even with
    monetary consequences,
    they suffer from multiple security issues.
    Many of these issues could be mitigated via an effective access control
    system, however, its realization is challenging due to
    the properties of current blockchain platforms (like lack of privacy, costly
    on-chain resources, or latency).
    To address this problem, we propose the \smacs framework, where updatable and
    sophisticated \textit{Access Control Rules (\acr)} for smart contracts can be realized with
    low cost. 
    \smacs shifts the burden of expensive \acr validation and management
    operations to an off-chain infrastructure, while implementing on-chain only lightweight
    token-based access control.
    \smacs is flexible and in addition to simple access control lists can easily implement
    rules enhancing the \textit{\runtime} security of smart contracts. 
    With dedicated \acr backed by
    vulnerability-detection tools, \smacs can protect vulnerable contracts 
    \textit{after deployment}. 
    We fully implement \smacs
    and evaluate it. 

\end{abstract}


\maketitle

\section{Introduction}
\label{sec:intro}
\input{sec/intro}

\section{Background and Motivation}
\label{sec:background}
\input{sec/background}

\section{SMACS Overview}
\label{sec:overview}
\input{sec/overview}

\section{SMACS Details}
\label{sec:details}
\input{sec/details}

\section{Runtime Verification Case Studies}
\label{sec:case}
\input{sec/case}

\section{Implementation and Evaluation}
\label{sec:evaluation}
\input{sec/evaluation}

\section{Discussion}
\label{sec:disc}
\input{sec/disc}

\section{Related Work}
\label{sec:related}
\input{sec/related}

\section{Conclusions}
\label{sec:conclusions}
\input{sec/conclusions}

\section*{Acknowledgment}
We thank the anonymous reviewers and our shepherd Yinzhi Cao 
for their valuable comments and suggestions.
This research is supported by the Ministry of Education, Singapore, under its
MOE AcRF Tier 2 grant (MOE2018-T2-1-111) and by the SUTD SRG ISTD 2017 128
grant. 

\bibliographystyle{IEEEtran}
\bibliography{paper}
\end{document}

%% file: code/solidity-highlighting.tex
\usepackage{listings, xcolor}

\definecolor{verylightgray}{rgb}{.97,.97,.97}

\lstdefinelanguage{Solidity}{
	keywords=[1]{anonymous, assembly, assert, balance, break, call, callcode, case, catch, class, constant, continue, constructor, contract, debugger, default, delegatecall, delete, do, else, emit, event, experimental, export, external, false, finally, for, function, gas, if, implements, import, in, indexed, instanceof, interface, internal, is, length, library, log0, log1, log2, log3, log4, memory, modifier, new, payable, pragma, private, protected, public, pure, push, require, return, returns, revert, selfdestruct, send, solidity, storage, struct, suicide, super, switch, then, this, throw, transfer, true, try, typeof, using, value, view, while, with, addmod, ecrecover, keccak256, mulmod, ripemd160, sha256, sha3}, 
	keywordstyle=[1]\color{blue}\bfseries,
	keywords=[2]{address, bool, byte, bytes, bytes1, bytes2, bytes3, bytes4, bytes5, bytes6, bytes7, bytes8, bytes9, bytes10, bytes11, bytes12, bytes13, bytes14, bytes15, bytes16, bytes17, bytes18, bytes19, bytes20, bytes21, bytes22, bytes23, bytes24, bytes25, bytes26, bytes27, bytes28, bytes29, bytes30, bytes31, bytes32, enum, int, int8, int16, int24, int32, int40, int48, int56, int64, int72, int80, int88, int96, int104, int112, int120, int128, int136, int144, int152, int160, int168, int176, int184, int192, int200, int208, int216, int224, int232, int240, int248, int256, mapping, string, uint, uint8, uint16, uint24, uint32, uint40, uint48, uint56, uint64, uint72, uint80, uint88, uint96, uint104, uint112, uint120, uint128, uint136, uint144, uint152, uint160, uint168, uint176, uint184, uint192, uint200, uint208, uint216, uint224, uint232, uint240, uint248, uint256, var, void, ether, finney, szabo, wei, days, hours, minutes, seconds, weeks, years},	
	keywordstyle=[2]\color{teal}\bfseries,
	keywords=[3]{block, blockhash, coinbase, difficulty, gaslimit, number, timestamp, msg, data, gas, sender, sig, value, now, tx, gasprice, origin},	
	keywordstyle=[3]\color{violet}\bfseries,
	identifierstyle=\color{black},
	sensitive=false,
	comment=[l]{//},
	morecomment=[s]{/*}{*/},
	commentstyle=\color{gray}\ttfamily,
	stringstyle=\color{red}\ttfamily,
	morestring=[b]',
	morestring=[b]"
}

%% file: sec/intro.tex
Blockchain-based platforms like Ethereum~\cite{buterin2017ethereum} have made
the concept of self-enforcing smart contract~\cite{szabo1996smart} into reality.
A smart contract is a special computer program that executes on the global
virtual machine running upon the 
distributed and
decentralized ledger. 
By running a consensus protocol and following the
replicated state machine model a unified view of the system state over all
network participants is imposed. 

Like all computer programs, it is likely that most non-trivial smart
contracts will contain
errors~\cite{Rubixi,HackerGold,daoatt}. 
These smart contract related errors should be addressed
even more seriously than ordinary program bugs. 
Firstly, smart contracts are often deployed over transparent and permissionless
blockchain platforms, thus anyone  
can inspect and interact with them.
Secondly, due to its immutability, it is hard to
upgrade or simply "kill" an already-deployed smart contract when attacks are
discovered
as the contract could have become an important part of the ecosystem
(i.e., other contracts hardcode its address).
Finally, smart contracts determine how units of value convertible
to real money move, making them a high-value target with intrinsic
economic incentives.
In the past few years, several hundreds of millions worth of 
USD were stolen or frozen due to flawed smart contracts~\cite{atzei2017survey,brent2018vandal}.
For instance, the infamous attack on the TheDAO~\cite{daoatt} smart contract resulted in over 50 million US Dollars worth of Ether were drained at the time the attack occurred. Given the severity of the attack, the Ethereum community finally agreed on hard-forking.

As a consequence, the community has made a great effort on
developing methodologies and tools to ensure the security of
smart contracts. 
One line of research focuses on the
security analysis of smart contracts by verifying their
code~\cite{luu2016making,Manticore,Mythril,tsankov2018securify}.
However, most of these approaches are unable to protect deployed smart contracts.
Another approach is to integrate \runtime defensive 
mechanisms into the deployed smart contracts and their \runtime environment.
With this approach, security-risky interactions with a vulnerable smart contract 
can be detected and mitigated ``on-the-fly'' in \runtime~\cite{grossman2017online,rodler2018sereum}.
These mechanisms usually require integration with the execution environment
(i.e., with the virtual machine deployed) to
be useful in production, but unfortunately that hinders their adoption.
Ideally, a defensive mechanism with arbitrary complexity would be put into a
smart contract itself, but in practice it is infeasible since on-chain resources
are expensive. Moreover, such a mechanism would be difficult to manage and update,
\acr would be publicly visible, and
available smart contract languages with virtual machines associated would limit its capabilities. 
For example, it would
be very costly and inconvenient to enhance the security of smart contracts by
encoding fine-grained \acr into them, which is a fairly
mature and traditional approach for centralized systems.

In this work, we propose the \smacs framework, a cost-effective access control service 
that is not simply a token-based authentication system but aims to enhance the 
\textit{\runtime security} of smart contracts.
In \smacs, a smart contract only needs to perform lightweight token verifications,
which introduces a clear on/off-chain sides separation with minimized on-chain
trusted computing base.
The on-chain storage and computation requirement are minimized by various techniques and by
shifting the burden of access control and ACR management to an off-chain
service. 
\smacs framework not only supports fine-grained and updatable 
\acr, but also is easily extensible by integrating recently 
developed smart contract vulnerability detection tools and security enhancement mechanisms~\cite{breidenbach2018enter,grossman2017online}.
The combination of SMACS with \runtime verification tools is powerful
as it provides the benefits of these tools immediately without requiring updating
virtual machines by all blockchain participants.
\smacs is deployable as today, does not require any blockchain platform changes,
and by moving security rules off-chain preserves their privacy.

%% file: sec/background.tex
\subsection{Blockchain and Smart Contracts}
In the last ten years, the blockchain technology initiated by 
the Bitcoin~\cite{underwood2016blockchain} cryptocurrency has fueled great innovations. 
Relying on the consensus protocol behind Bitcoin,
it is not long before Ethereum~\cite{buterin2017ethereum} was built -- 
a general-purpose blockchain system
well-beyond cryptocurrencies which can execute
programs on blockchain.  
Similarly to Bitcoin, Ethereum introduces its native cryptocurrency --
\textit{ether}, which is also used to incentivize system participants.

Ethereum can be regarded as a decentralized and
replicated state machine whose state
is maintained as a proof-of-work blockchain.
The state transition of Ethereum is processed by the so-called
\textit{Ethereum Virtual Machine}~(EVM) 
executing programs called smart contracts
written in a Turing-complete language.
Due to the Turing completeness, 
one can implement self-enforcing smart contracts with nearly arbitrary logic.
As a result, we have witnessed a wide range of applications of smart contracts in
different domains~\cite{breidenbach2018enter,korpela2017digital}.

Smart contracts inherit some essential properties from
its underlying blockchain. 
Once a smart contract is deployed on the Ethereum 
platform, its code is immutable and visible by every node in the Ethereum network, and all
transactions calling it are also transparent to all. 
Therefore, it is the duty of the contract developer to implement 
proper access control or defensive mechanisms prior to deployment, and failed to do so 
can lead to tremendous and irreversible financial losses.

\subsection{EVM and Solidity}\label{sect:programming}
EVM executes smart contracts 
as a Turing complete stack-based language at low level called 
Ethereum bytecode. In practice, smart contracts are typically developed
in high-level languages such as Solidity, Vyper, Serpent, etc., and are
compiled into bytecode by the corresponding compilers.\footnote{Note that even a
smart contract is semantically bug-free with respect to the underlying
high-level language, compilers may introduce language-specific vulnerabilities
into the
system~\cite{atzei2017survey,hirai2016formal,amani2018towards,bhargavan2016formal,grishchenko2018foundations},
which once again highlight the importance of \runtime security analysis.}
In this work we focus on Solidity as it is the most popular and developed
language
of Ethereum.
There are three memory regions of a smart contract program: stack, memory, and storage. The \textit{stack} and \textit{memory}
are volatile and cheap to use. The \textit{storage} is maintained on blockchain 
and is the only persistent memory
region across transactions.

The execution logic of a smart contract is modularized into
methods which are the executable code segments within a contract~\cite{mohanty2018basic}. 
Method calls can happen internally or externally and have different levels of 
visibility towards other contracts. There are four types of visibilities for
methods in Solidity:

\setdefaultleftmargin{0em}{0em}{}{}{}{}
\begin{compactitem}
\item \textbf{external} methods are part of the contract interface, which means
    they can be called from other contracts and via transactions.
     An external methods cannot be called internally, 

\item \textbf{public} methods are part of the contract interface and can be
    either called internally or via messages (see the next section),

\item \textbf{internal} methods can only be accessed internally (i.e., from
    within the current contract or contracts deriving from it),

\item \textbf{private} methods are only visible for the contract they are defined in and not in derived contracts.
\end{compactitem}

All computational or memory utilization
in Ethereum is charged in \textit{gas}, which can be
regarded as a separate virtual currency with its own 
exchange rate against ether~\cite{zhang2016town}.
The gas system is essential to incentivize system
participants and prevent
denial-of-service attacks or inadvertently resource-devouring transactions.
Performing 
computation or storing 
data objects of large size (e.g., access rules) can 
be gas-expensive.  
In fact, according to our experiment, creating even a simple whitelist with 10k addresses 
would cost around \$300. Also managing such a list would have a linear cost in the number of 
update operations.

\subsection{Transactions and Message calls}\label{subsec:txcall}
In Ethereum, every state change of the global singleton state machine 
is ultimately due to a \textit{transaction}, a signed data package originated
from an externally owned account. Transactions are recorded on the
blockchain and can move value from
one account to another or/and trigger smart contract execution. 
User accounts and smart contract instances are uniformly identified by unique addresses.

Contracts can call other contracts or send ether to non-contract accounts by 
\textit{message calls}, the virtual objects that are never serialized 
and exist only in the Ethereum execution environment. Every transaction 
consists of a top-level message call which in turn can create further message calls.
It implies that
from a simple transaction initiated by an externally owned account, a \textit{call chain} of contract executions can be triggered. 
Solidity allows smart contracts to access some global objects and properties of
the blockchain~\cite{global}. In the context of this work the following objects are relevant: 
\begin{compactitem}	
	\item \textbf{tx.origin} - the sender of the transaction for full call chain (a list of all called methods that a given transaction triggers),

    \item \textbf{msg.sender} - the sender of the message for the current call. 
        Let us consider a call chain triggered by 
        a transaction $T$ originated from the externally owned
        account $u$, where $T$ calls the contract $A$, and $A$ calls another contract $B$. 
        Then from $B$'s perspective
        the value of \texttt{msg.sender} is the address of $A$ while 
        \texttt{tx.origin} is the address of $u$,

    \item \textbf{msg.sig} - the method identifier (encoded as the first four
        bytes of \textit{calldata}). Ethereum tags identifiers of each method
        for every smart contract. When $A$ calls \texttt{B.funcB(argA,argB)},
        the value of \texttt{msg.sig} seen by $B$ would be the identifier of
        \texttt{funcB()},

    \item \textbf{msg.data} - the complete calldata (the method identifier and
        passed arguments). The value of \texttt{msg.data} for the case
        above is \texttt{msg.sig} appended with the encoded values of
        \texttt{argA} and \texttt{argB}. 
\end{compactitem}

Transactions are signed by their originators and before processing them in EVM
their authenticity is validated by the Ethereum network. 
To prevent replay attacks each transaction has
a nonce which also is validated by participants. 
However, nonces
cannot be accessed by Solidity contracts.

\subsection{Access Control in Smart Contracts}
Access control is a security technique that
regulates
who has access to certain system resources. 
The intention of access control in Ethereum smart contracts is to
restrict the access to contract functionalities according to suitable criteria.
In~\autoref{sect:programming}, the programming language
of smart contracts already has some features to facilitate a minimum level
of access control. However, these features are limited, only defining access control rules for built-in 
methods. 

In general, implementing access control for smart contracts 
based on a permissionless blockchain is difficult. 
A naive approach of putting all access control logic into
a smart contract would 
violate privacy and consume a lot of expensive on-chain resources.
On the other hand, due to the immutability 
nature of the underlying blockchain, it is difficult to update the \acr if they are managed over blockchain. 
To the best of our knowledge,
there is no generic framework in the literature realizing
efficient and flexible access control for smart contracts.
However, such a framework could be highly beneficial for the security of smart
contracts and the ecosystem.
Currently, 
many token sales allow only approved users to participate in a token
sale or trade. To achieve \textit{on-chain} access control, the owner of the token
sale contract maintains a whitelist listing addresses of authorized users.  
When a user tries to access the contract, there is an access control check verifying
whether the user is whitelisted. 
For example, the Bluzelle decentralized
database has paid 9.345 ETH (11,949 USD at the time) just to
whitelist 7473 users for their token sale~\cite{Bluzellepaid}.
Similarly, OpenZeppelin~\cite{OpenZeppelin} provides templates like role-based 
access control or ``proxy contracts''.
Unfortunately, these solutions are intended for on-chain contract
management,
do not allow for flexible changes at runtime,
and still have
all other limitations of on-chain access control.

To further illustrate the motivations behind our system, we give
several other examples with brief comments.
\begin{example}\label{eg:whitelist}
A service provider may want to create a smart contract whose methods can be
called only by a dynamic set of addresses (e.g., employees or business
partners).  	
\end{example}

\begin{example}\label{eg:blacklist}
The owner of a contract may want her smart contract to block 
the access from a predefined set of addresses.  
\end{example}

~\autoref{eg:whitelist} and~\autoref{eg:blacklist} show the needs of
implementing very basic and common \acr such as whitelist
and blacklist. Note that the involved lists should be updatable dynamically,
and the method for defining the lists should be flexible enough.
Implementing even such a basic access control on-chain in smart contracts would be highly
expensive if a black/whitelist is long and/or updated frequently.
Moreover, managing such an access control list would be impractical as executing blockchain
transactions is significantly delayed (minutes to hours~\cite{weber2017availability}).
The list maintained on-chain would be also visible to anyone which in most cases is undesired.

\begin{example}
The owner of a contract may require that only authorized parties can
call a specific method. She may demand more fine-tuned controls:
only authorized parties can call the method with specific arguments.
\end{example}

This kind of selective restrictions may be useful in almost any smart contract
application, e.g., to determine who can move the money, 
who can stop a service, etc. The latter rule hints at an even more exciting
application of access control: \textit{the owner can specify that an address can
call a method in a smart contract only when the payload will not trigger any
known security problems}. By extending this kind of rules, we may prevent attacks on
vulnerable contracts even \textit{after their deployment} -- see the next example.

\begin{example}
The owner of a contract may require that a given call can be executed only when
it is validated by some sophisticated \runtime verification tool(s).
\end{example}

In contrast to the 
previous 
examples, such a rule
would allow the owner to inspect a given call in detail and limit access in the
case of any issue detected. In such a way the owner could benefit from \runtime
verification tools running them off-chain, without 
integrating them in
the smart contract environment.
The owner may also want to ensure a given call can be executed only once, 
unless a new permission is granted.

%% file: sec/overview.tex
Throughout the paper, we describe \smacs in the context of Ethereum and Solidity, but it can be
easily extended to other platforms and languages with similar capabilities.

\subsection{System Participants}
\label{subsect:participants}
\smacs involves four types of actors:

\begin{compactitem}
\item \textbf{\smacs-enabled Smart Contracts} are contracts 
        on blockchain protected by \smacs. 
        A \smacs-enabled smart contract
        verifies incoming calls
        by validating their corresponding tokens.  Any transaction or a message
        call will be rejected if a valid token is not presented.  
        For simple description, we often refer to \smacs-enabled smart contracts as ``smart
        contracts'' or just ``contracts'' and from the context it will be clear
        when we distinguish them from legacy smart contracts.

\item \textbf{Owner} is the creator of a smart contract. 
Normally, there could be several smart contracts under the control 
of a single owner. An owner is responsible for defining and managing the \acr. Also, an owner needs to manage a \ts (TS) 
instance corresponding to a \smacs-enabled smart contract.
\item \textbf{\ts (TS)} is a service that is responsible for verifying
    requests from clients and issuing access control \textit{tokens}
        accordingly. A token issued by a TS determines exact access
        permissions of a particular client with respect to a \smacs-enabled smart
        contract. 
\item \textbf{Clients} 
are users who want to access the resources (e.g., data, methods) of
\smacs-enabled smart contracts. A client must obtain a token granting
appropriate permissions from the TS before she can access the smart contracts.
\end{compactitem}

\subsection{Goals}
\label{sect:goals}
We design \smacs with the following goals in mind.
\paragraph{\textbf{Security}}
We assume that an adversary cannot compromise 
the underlying cryptographic primitives (e.g., signatures, hash functions,
etc.) and cannot compromise the \runtime environment of the deployed smart contract platform 
However, we discuss an adversary able to reverse the blockchain history (i.e., launch a \textit{51\% attack}).
Under these assumptions \smacs should prevent unauthorized entities from accessing 
\smacs-enabled smart contracts. When an entity is allowed to
access a \smacs-enabled smart contract, its behavior cannot
deviate from the permission it has been granted. 
Moreover, apart from enabling access control in the traditional sense,
\smacs should be able to counter certain \runtime
attacks even if the underlying smart contracts are vulnerable to 
these attacks. 

\paragraph{\textbf{Flexibility and  Extensibility}}
The \smacs framework should be able to define complex and fine-grained 
\acr for smart contracts
while keeping smart contracts simple.
\smacs should allow to manage these rules by removing, adding, or modifying
them dynamically, but without updating contracts.
Also, it should be easy to extend \smacs by integrating 
smart contract protection techniques of various classes.

\paragraph{\textbf{Efficiency and Low cost}}
\smacs should operate as efficiently as possible. There should
be no efficiency bottlenecks with respect to throughput, storage,
latency, etc. which could hinder its applicability in real-world scenarios.
The cost of applying \smacs in terms of storage, computation,
and blockchain-related fees should be minimized. 
Meanwhile, the process of integrating and
deploying \smacs-enabled smart contracts should be easy and intuitive, not
incurring a high development effort and cost.

\begin{figure}[t!]
\setlength{\belowcaptionskip}{-0.2cm}
    \centering
    \includegraphics[width=0.9\linewidth]{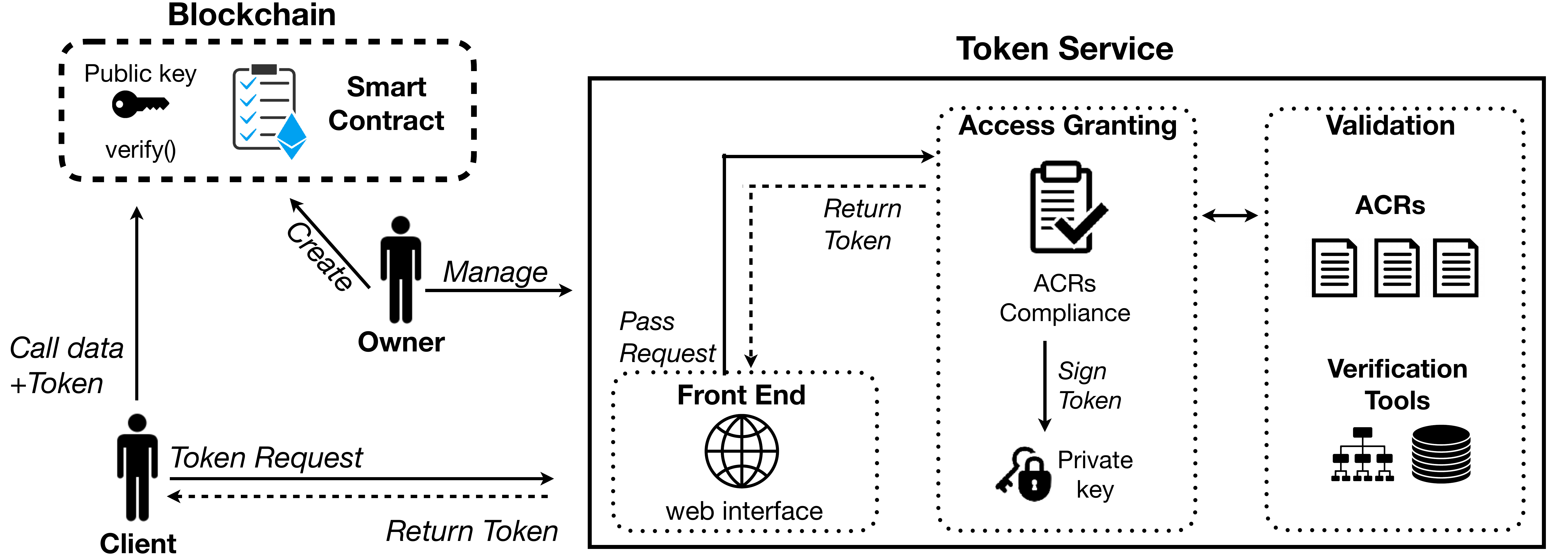}
    \caption{Details of the \smacs framework.}
    \label{fig:imple}
\end{figure}

\subsection{Overview}
\label{subsect:interactions}

In \smacs, the owner first generates a public and private key pair 
$(pk_{TS}, sk_{TS})$, and 
preloads the \ts (TS) with $sk_{TS}$ and an initial set
of \acr (or token issuing rules).
The private key $sk_{TS}$ will be used by the TS
to sign tokens it issues, while the \acr
specify the condition under which a token can be issued. 
The owner is also responsible for creating the
\smacs-enabled smart contract with the public key $pk_{TS}$ preloaded. 
The \smacs-enabled smart contract verifies the validity of the access credentials (tokens) 
of the incoming calls with the public key $pk_{TS}$ before continuing an actual execution. 
We use ${Sign}_{sk}(\cdot)$ and ${SigVerify}_{pk}(\cdot)$ to denote the 
signing with $sk$ and verifying with $pk$ in
later sections.

Although tokens in \smacs can have different types and can be issued basing on
arbitrary validation logic, \smacs-enabled smart contracts stay simple and
implement only an easy access control verification.  In most cases, the only
overheads that \smacs introduce are a) storing a public key,
b) parsing a token, and c) a signature verification per call.  
In our design, the burden
of all memory consuming and computationally heavy operations are shifted to an
off-chain TS. 
All intended \acr are
initialized into TS. These rules can be updated dynamically by the owner.
Before accessing the \smacs-enabled smart contract, a client must apply for a
token with compatible permissions from TS. 
Upon receiving the request from a client, the TS checks the request against the
\acr. If the request does not violate the rules, the TS 
issues a token to the client by signing the datagram formed by relevant 
information extracted from the request and metadata. The client then constructs a transaction
with the token encoded into it to access the smart contract. 
The transaction constructed by the client has to be compatible with her previous
request, since the signature creates a cryptographic binding and any attempt to
modify the actual transaction will make the signature verification fail.  
In practice, the situation may be more complex. A client may 
issue a transaction that triggers the execution of a chain of smart contracts. 
How to handle these situations will be clear
in the following sections.

%% file: sec/details.tex
The detailed architecture of \smacs is shown in~\autoref{fig:imple}.  The owner
and clients are external owned accounts operated via the client-side software
(usually called a \textit{wallet}) to interact with \smacs-enabled
contracts. The TS consists of a) the \textit{front end} interface, b) the
\textit{access granting} module that checks the rules compliance and issues
tokens, and c) the \textit{validation} module that contains all verification
tools (if any) and respective rules.  
These rules determine who can get a token with particular permissions.  
The owner and clients interact with the TS through an
HTTPS-enabled web interface provided by the TS. 
The realization of the access control of the smart contract is ultimately 
due to the control of the issuance of
tokens.

\subsection{Token Types}
\label{subsect:semantics}
\smacs supports three different types 
of tokens with different permission semantics. These types 
are designed to facilitate a flexible and fine-grained access control over
the \smacs-enabled smart contracts. 

\begin{compactitem}
    \item \textbf{Super token} is of the highest permission level. 
    A client with a super token
    can freely call all public methods of the smart contract 
    with arbitrary arguments before the token expires. 
    
    \item \textbf{Method token} limits the access to a specific method.
    A client with a method token can call the specific contract's method
        associated with the token with arbitrary arguments before the token
        expires. A method token issued for a particular method cannot be used to
        access other methods.   
    
    \item \textbf{Argument token} is similar to a method token with the additional
        restriction that the associated method can only be called with specific
        arguments.  
\end{compactitem}

All tokens are issued with an expiration time set by the TS. The expiration
time determines the token lifetime, i.e., until when the token can be used to
authorize the corresponding calls.
Any token can have the \textit{one-time} property set. 
A one-time token gets invalidated once it is used to successfully access the
smart contract, 
which ensures that the token holder can only access the smart
contract once with issued token. 
\begin{figure}[!b]
\setlength{\belowcaptionskip}{-0.7cm} 
\setlength{\abovecaptionskip}{0.1cm}
    \begin{center}
        \begin{tikzpicture}[scale = 0.9,>=stealth]
        
        \draw (0, 0) rectangle node {\scriptsize $\texttt{type}$} +(0.9, 0.5);  
        \node[scale = 0.7] at (0.45, 0.45+0.25) {1B};

        \draw (0.9, 0) rectangle node {\scriptsize $\texttt{cAddr}$} +(0.9, 0.5);  
        \node[scale = 0.7] at (1.35, 0.45+0.25) {20B};
        
        \draw (1.8, 0) rectangle node {\scriptsize $\texttt{sAddr}$} +(0.9, 0.5);  
        \node[scale = 0.7] at (2.25, 0.45+0.25) {20B};
        
        \draw (2.7, 0) rectangle node {\scriptsize $\texttt{methodId}$} +(1.7, 0.5);
        \node[scale = 0.7] at (3.55, 0.45+0.25) {string};
        
        \draw (4.4, 0) rectangle node {\scriptsize $\texttt{argName}$} +(1.5, 0.5);
        \node[scale = 0.7] at (5.15, 0.45+0.25) {string};
        
        \draw (5.9, 0) rectangle node {\scriptsize $\texttt{argValue}$} +(1.5, 0.5);
        \node[scale = 0.7] at (6.65, 0.45+0.25) {string};
        
        \draw (7.4, 0) rectangle node {\scriptsize $\texttt{\ldots}$} +(0.6, 0.5);

        \draw[dotted] (0,0) -- (0, -0.5);
        \draw[dotted] (8,0) -- (8, -0.5);

        \node[scale = 0.7] at (4.45, -0.25) {\texttt{reqPayload}};
        \draw[->] (4.45-0.9, -0.25) -- (0.9, -0.25);
        \draw[->] (4.45+0.9, -0.25) -- (8, -0.25);
        
        \end{tikzpicture}
        \caption{The layout of a token request.}
        \label{fig:request}
    \end{center}
\end{figure}
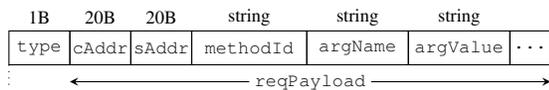

Before a client can get a token from the TS, 
she has to submit a well-formed \textit{token request} to the TS.
As depicted in~\autoref{fig:request}, 
the token request varies according to the requested token type, 
where 
\texttt{type} is a token type,
\texttt{cAddr} is the address of a targeted contract,
\texttt{sAddr} is the address of a client's account,
\texttt{methodId} is the method identifier that is going to be accessed with a method or argument token,
\texttt{argName} and \texttt{argValue} are the argument and argument value
used when an argument token is requested (there can be multiple argument-value
pairs passed in a token request).

In \smacs, a token is implemented as an $86$-byte object shown in~\autoref{fig:token}, where 
\texttt{type} indicates the type of a token, \texttt{expire} encodes
the expiration time, \texttt{index} is used for tokens with their one-time property set
(if the value of \texttt{index} is a non-negative integer, then the one-time property is set),
and the \texttt{signature} field is computed as: 
$$
\texttt{Sign}_{sk_{TS}}({\color{blue} \texttt{type}}  \parallel \texttt{expire} \parallel \texttt{index} 
\parallel {\color{blue}\texttt{reqPayload}} ),
$$ 
where
${\color{blue}\texttt{type}}$ and ${\color{blue}\texttt{reqPayload}}$ 
are extracted from the token request sent by the client 
(see~\autoref{fig:request}).
The ${\color{blue}\texttt{reqPayload}}$ is an optional 
field of the token request with variable size according 
to ${\color{blue}\texttt{type}}$. Its exact
formulation is shown in~\autoref{table:atomictoken}.

\begin{table}[t!]
\setlength{\belowcaptionskip}{0.1cm}
\setlength{\abovecaptionskip}{0.1cm}
    \caption{Elements of a token request payload.}
    \label{table:atomictoken}\centering
    \begin{tabular}{L{1.3cm}C{0.65cm}C{0.65cm}C{1.2cm}C{1.0cm}C{1.2cm}}
        \toprule
        \multirow{2}*{Type} & \multicolumn{5}{c}{\texttt{reqPayload}}\\ \cmidrule{2-6}
               &\texttt{cAddr} &\texttt{sAddr}  &\texttt{methodId} &\texttt{argName} &\texttt{argValue}\\ \midrule
        Super       &\cmark &\cmark     &\xmark     &\xmark  &\xmark    \\
        Method      &\cmark &\cmark     &\cmark     &\xmark  &\xmark            \\
        Argument    &\cmark &\cmark     &\cmark     &\cmark  &\cmark     \\ \bottomrule
    \end{tabular}
\end{table}

\begin{figure}[!ht]
\setlength{\belowcaptionskip}{-0.4cm}
\setlength{\abovecaptionskip}{0.1cm}
    \begin{center}
        \begin{tikzpicture}[scale = 0.8,>=stealth]
        \draw (0, 0) rectangle node {\scriptsize $\texttt{type}$} +(1, 0.5);  
        \node[scale = 0.7] at (0.5, 0.5+0.25) {1B};
        
        \draw (1, 0) rectangle node {\scriptsize $\texttt{expire}$} +(2-0.5, 0.5);
        \node[scale = 0.7] at (1.75, 0.5+0.25) {4B};
        
        \draw (2.5, 0) rectangle node {\scriptsize $\texttt{index}$} +(2.5, 0.5);
        \node[scale = 0.7] at (3.75, 0.5+0.25) {16B};
        
        \draw (5, 0) rectangle node {\scriptsize $\texttt{signature}$} +(3, 0.5);
        \node[scale = 0.7] at (6.5, 0.5+0.25) {65B};
        
        \draw[dotted] (0,0) -- (0, -0.5);
        \draw[dotted] (8,0) -- (8, -0.5);
        
        \node[scale = 0.7] at (4, -0.25) {86 bytes};
        \draw[->] (4-1, -0.25) -- (0, -0.25);
        \draw[->] (4+1, -0.25) -- (8, -0.25);
        
        \end{tikzpicture}
        \caption{The layout of a token.}
        \label{fig:token}
    \end{center}
\end{figure}
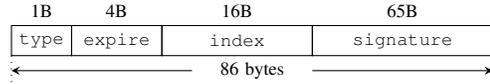

\subsection{Token Issuance and Verification}
\label{subsec:token_issue_verify}
There are two verification processes in \smacs:
a TS verifies incoming token requests against its rules, 
and a \smacs-enabled contract verifies tokens extracted from incoming transactions.
Any failed attempt to access the contract is ultimately
due to the failure of one of these two verification processes.

\paragraph{\textbf{Token Issuance}}
To apply for a token, a client sends a token request 
specifying the intended \texttt{type} together with 
a compatible \texttt{reqPayload}, which
describes who (\texttt{sAddr}) will access which (\texttt{cAddr}) 
smart contract and how it will be accessed. 
The request payload (\texttt{reqPayload}) depends on the intended \texttt{type}
(see~\autoref{table:atomictoken}).
When receiving the token request, the TS parses and checks 
it against the rules. Once verified, a token is issued according to the request.
This step can be easily integrated into mainstream
wallets, such that it is executed seamlessly for users prior to actual transaction sending.

\paragraph{\textbf{Contract-side Verification}}
Once getting a token from the TS, the client 
can construct a transaction whose calldata is 
filled with the token together with other necessary 
information that is compatible with the token
(i.e., the token will be passed as an argument).
Upon receiving the transaction, the token verification process
shown in~\autoref{alg:single} is triggered, and only after
the process succeeds the smart contract can continue to
execute the transaction accordingly.

The \smacs-enabled smart contract first extracts 
the token from the transaction,
checks whether it expires, and whether it has been used 
if the one-time property 
is set.
We discuss the one-time property validity checking (i.e., the details of $reused()$ check)
in~\autoref{subsect:onetime}.
Then the \smacs-enabled smart contract reconstructs the data required to verify
the token signature according to the type of the token.  
In this step, the smart
contract uses the EVM's transaction context objects (
see~\autoref{subsec:txcall}) to make sure that the passed
ticket matches the current transaction.

\begin{algorithm}[b!]
  \caption{Contract-side token verification.}\label{alg:single}
  \small
  \DontPrintSemicolon
  \KwIn{A transaction $T$}
  \KwOut{The verification result}
  $tk \leftarrow \texttt{extractToken}(T)$\\
   
  \If{$now() > tk\texttt{.expire}$}
  {\Return~~{\it False}}

    \If{$tk.\texttt{index} > -1$~$\mathrm{and~not}$~$reused(tk.\texttt{index})$}
  {\Return~~{\it False}}
  
  $tkData \gets tk.\texttt{expire} \parallel tk.\texttt{index}$\\
  $addrData \gets  {T\texttt{.origin}} \parallel {address(this)} $\\
  $data \gets tk.\texttt{type} \parallel tkData \parallel addrData$\\

  \uIf{$tk\texttt{.type} = \mathrm{Super}$}{$data \gets data$\\}
  \uElseIf{$tk\texttt{.type} = \mathrm{Method}$}{$data \gets data \parallel {msg.sig}$\\}
  \ElseIf {$tk\texttt{.type} = \mathrm{Argument}$}{$data \gets data \parallel {msg.sig} \parallel {msg.data} $\\}
  \Return~~ $\texttt{SigVerify}_{pk_{TS}}(data, tk\texttt{.signature})$\\
\end{algorithm}

We emphasize that this verification process is generic
and the implementation of \smacs-enabled smart contracts
respects the common development flows.
Basically, the code of legacy smart contracts 
can be made deployment-ready in the \smacs framework by 
ensuring that every method callable from outside (i.e., public or external)
verifies a token prior to its actual body execution.
To achieve it, for each public and external method the 
\texttt{tokens} argument is added to the original argument
list and the \texttt{verify} call performing~\autoref{alg:single} 
is asserted prior to the actual method body.

\begin{figure}[!t]
\setlength{\belowcaptionskip}{-0.3cm}
    \centering
    \includegraphics[width=0.8\linewidth]{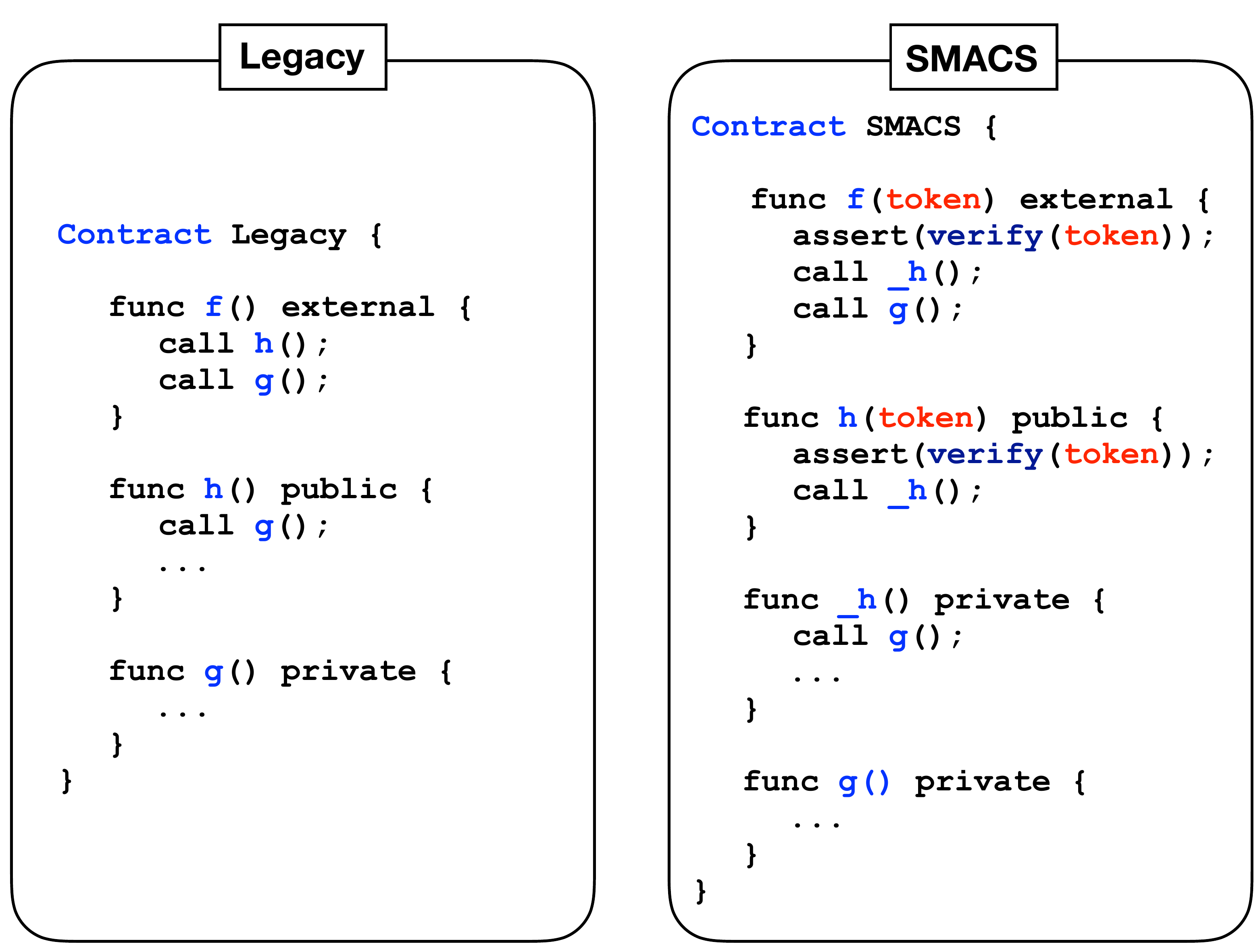}
    \caption{Automated \smacs adoption for a legacy contract.}
    \label{fig:interface}
\end{figure}

To facilitate easy adoption
we develop a tool that allows to transform any legacy smart contract into
an equivalent \smacs-enabled smart contract.  
An example of such a transformation is presented
in~\autoref{fig:interface} (note that internal methods do not have to verify
tokens and in the case of public/external methods
called internally they are split into separate methods).

\subsection{One-time Tokens}
\label{subsect:onetime}
The one-time property ensures that a given token can be used only once.
One-time tokens 
may be especially useful for access control of security-critical
methods or for example when
a client is unknown to the requested TS.
To realize this property, one may be tempted to rely on the \textit{nonce} mechanism
employed by Ethereum for counteracting replay attacks.
However, we emphasize that the transaction's nonce cannot be accessed by the smart contract
itself. Therefore,
\smacs has to implement an in-contract mechanism to
support the verification of one-time tokens.

The TS maintains \texttt{counter} variable (corresponding to the \texttt{index} field of a token)
for issuing tokens with the one-time property set.
\texttt{counter} is initialized to $0$, 
whenever a new one-time token is being issued, it is incremented by 1, 
and the updated
valued is used as the \texttt{index} value for this token.

Since a one-time token is supposed to be used
only \textit{once}, when a client with such a token tries to access
an \smacs-enabled smart contract,
the contract has to check whether the underlying token has been 
used before, and then permits or denies the access attempt accordingly.
A trivial way for the contract to realize this is to 
store the \texttt{index} values of all one-time tokens
having made a successful access.
However, as the on-chain storage is expensive, 
this approach can be costly and impractical
if the number of issued one-time tokens is large. 

Based on the observations that the TS can assign 
every one-time token with a unique index consecutively
and the token lifetime is limited, we propose a cost-effective scheme
to handle one-time tokens, where every index is efficiently encoded as 
a single bit of a cyclically reused bitmap.
In our approach, an $n$-bit map $S$ (together with its internal state)
is used to represent the status (used or unused) of a set of $n$ one-time
tokens with consecutive \texttt{index} values.
The state of the bitmap can be represented by a tuple
\textit{(S, start, startPtr, end, endPtr)}, 
where $S\in \{0,1\}^n$, $ start \in \{0, 1, \cdots \}$, $startPtr 
\in \{ 0, \cdots, n-1 \}$, $end = start + n-1$, and $endPtr = startPtr + n-1 ~ \mathrm{mod}~ n$.
In~\autoref{alg:move}, the $n$-bit sequence
$
S\left[startPtr\right] \parallel  S\left[startPtr + 1~\mathrm{mod}~n\right] \parallel\cdots \parallel S\left[endPtr\right]
$
indicates the status of the $n$ one-time tokens with \texttt{index}es
$start, start+1, \cdots$, and $end$. A token with the index $i$ is regarded as
unused if and only if a) $i \in \{ start, start+1, \cdots, start + n-1 \}$
and $S[(startPtr + i-start)~\mathrm{mod}~n] = 0$, or b) $i > end$.
When the index $i$ of a token is unused, the state of the bitmap is updated
according to~\autoref{alg:move}. 
\begin{algorithm}[t!]
  \caption{Bitmap state update.}\label{alg:move}
  \DontPrintSemicolon
  \small

  \texttt{/* Initialization */}\\
  $S \leftarrow [0, \cdots, 0]$;
  $start \leftarrow 0$; 
  $end \leftarrow n-1$;\\
  $startPtr \leftarrow 0$;
  $endPtr \leftarrow n - 1$;\\
  \texttt{/* Update */}\\
  
        $i \leftarrow \texttt{getIndex}(\mathrm{token})$\\
        \uIf{$i < start$}
        {\Return False}
        \uElseIf{$ start \leq i \leq end$}
        {
            $t \gets (startPtr + i - start)~\mathrm{mod}~n$\\
            \uIf{$ S[t] = 1 $}
            {\Return False}
            \Else
            {
                $S[t] \gets 1$\\
                \Return True\\
            }
        }
        \uElseIf{$ end < i \leq end + n$ }
        { 
            $startPtr \gets \texttt{seek}(S, i, end, startPtr)$\\
            $endPtr \leftarrow (startPtr + n - 1) ~\mathrm{mod}~n$\\
            $end \leftarrow i$;
            $start \leftarrow end - n + 1$;
            $S[endPtr] \leftarrow 1$;\\
            \Return True\\
        }   
        \Else
        {   
            \texttt{/* Reset when $i$ is too large */}\\    
            $S \leftarrow [0, \cdots, 0]$;
            $startPtr \leftarrow 0$;
            $endPtr \leftarrow n-1$;\\
            $start \leftarrow i$;
            $end \leftarrow i + n - 1$;
            \Return True\\
        }
\vspace{2mm}
  \texttt{/* $\texttt{seek}(S, i, end, startPtr)$ returns
    the smallest integer $j$ in $\{0, \cdots, n-1\}$  such that $S[j] = 0$ and $i - end \leq j - startPtr$
     */}\\
\end{algorithm}

Let us consider a bitmap $S$ with size $n = 8$. 
At the beginning, all cells of $S$ are set to 0 with
$start = startPtr = 0$ and $end = endPtr = 7$. 
\begin{center}
    \begin{tikzpicture}[scale = 0.7,>=stealth]
    
    \draw (0, 0) rectangle node {\scriptsize $\texttt{0}$} +(1, 0.5);
    \draw (1, 0) rectangle node {\scriptsize $\texttt{1}$} +(1, 0.5);
    \draw (2, 0) rectangle node {\scriptsize $\texttt{2}$} +(1, 0.5);
    \draw (3, 0) rectangle node {\scriptsize $\texttt{3}$} +(1, 0.5);
    \draw (4, 0) rectangle node {\scriptsize $\texttt{4}$} +(1, 0.5);
    \draw (5, 0) rectangle node {\scriptsize $\texttt{5}$} +(1, 0.5);
    \draw (6, 0) rectangle node {\scriptsize $\texttt{6}$} +(1, 0.5);
    \draw (7, 0) rectangle node {\scriptsize $\texttt{7}$} +(1, 0.5);
    
    \draw[->, thick] (0.5, 1-0.1) -- (0.5, 0.5);
    \node[scale = 0.8] at (0.5, 1+0.1) {$start: 0$};
    
    \draw[->, thick] (7+0.5, 1-0.1) -- (7+0.5, 0.5);
    \node[scale = 0.8] at (7+0.5, 1+0.1) {$end: 7$};
    \end{tikzpicture}
\end{center}
After tokens whose \texttt{index} are $0$, $1$, $4$, and $5$ access
the smart contract, the corresponding cells are set to 1 
(gray cells).
\begin{center}

    \begin{tikzpicture}[scale = 0.7,>=stealth]
    
    \fill[color = lightgray] (0, 0) rectangle +(1, 0.5);
    \fill[color = lightgray] (1, 0) rectangle +(1, 0.5);
    \fill[color = lightgray] (4, 0) rectangle +(1, 0.5);
    \fill[color = lightgray] (5, 0) rectangle +(1, 0.5);
    
    \draw (0, 0) rectangle node {\scriptsize $\texttt{0}$} +(1, 0.5);
    
    \draw (1, 0) rectangle node {\scriptsize $\texttt{1}$} +(1, 0.5);
    \draw (2, 0) rectangle node {\scriptsize $\texttt{2}$} +(1, 0.5);
    \draw (3, 0) rectangle node {\scriptsize $\texttt{3}$} +(1, 0.5);
    \draw (4, 0) rectangle node {\scriptsize $\texttt{4}$} +(1, 0.5);
    \draw (5, 0) rectangle node {\scriptsize $\texttt{5}$} +(1, 0.5);
    \draw (6, 0) rectangle node {\scriptsize $\texttt{6}$} +(1, 0.5);
    \draw (7, 0) rectangle node {\scriptsize $\texttt{7}$} +(1, 0.5);
    
    \draw[->, thick] (0.5, 1-0.1) -- (0.5, 0.5);
    \node[scale = 0.8] at (0.5, 1+0.1) {$start: 0$};
    
    \draw[->, thick] (7+0.5, 1-0.1) -- (7+0.5, 0.5);
    \node[scale = 0.8] at (7+0.5, 1+0.1) {$end: 7$};
    \end{tikzpicture}
\end{center}
Upon receiving the access request by the token with \texttt{index}
$9$, $\texttt{seek}()$ is responsible for finding the updated 
$startPtr$. Since $  end = 7 < 9 \leq end + 8 = 15 $, $\texttt{seek}()$
returns $2$, which is assigned to $startPtr$, and $S[endPtr]$ is set to $1$,
where $endPtr \leftarrow startPtr + n-1 ~\mathrm{mod}~n = 2 + 8 - 1 ~\mathrm{mod}~8 = 1$.

\begin{center}

    \begin{tikzpicture}[scale = 0.7,>=stealth]
    \fill[color = lightgray] (1, 0) rectangle +(1, 0.5);
    \fill[color = lightgray] (4, 0) rectangle +(1, 0.5);
    \fill[color = lightgray] (5, 0) rectangle +(1, 0.5);
    \draw (0, 0) rectangle node {\scriptsize $\texttt{0}$} +(1, 0.5);
    \draw (1, 0) rectangle node {\scriptsize $\texttt{1}$} +(1, 0.5);
    \draw (2, 0) rectangle node {\scriptsize $\texttt{2}$} +(1, 0.5);
    \draw (3, 0) rectangle node {\scriptsize $\texttt{3}$} +(1, 0.5);
    \draw (4, 0) rectangle node {\scriptsize $\texttt{4}$} +(1, 0.5);
    \draw (5, 0) rectangle node {\scriptsize $\texttt{5}$} +(1, 0.5);
    \draw (6, 0) rectangle node {\scriptsize $\texttt{6}$} +(1, 0.5);
    \draw (7, 0) rectangle node {\scriptsize $\texttt{7}$} +(1, 0.5);
    \draw[->, thick] (2+0.5, 1-0.1) -- (2+0.5, 0.5);
    \node[scale = 0.8] at (2+0.5+0.2, 1+0.1) {$start: 2$};
    \draw[->, thick] (1+0.5, 1-0.1) -- (1+0.5, 0.5);
    \node[scale = 0.8] at (1+0.5-0.2, 1+0.1) {$end: 9$};
    \end{tikzpicture}
\end{center}
At this point, 
$S[2]\parallel S[3] \parallel \cdots \parallel S[7] \parallel S[0] \parallel S[1] $
represents the status of the tokens with \texttt{index}es in $\{2, 3, \cdots, 9 \}$.
We continue to consider a more complicated case where an access request is made
by the token with \texttt{index} $13$. Then the state of the bitmap is updated
as follows. 
\begin{center}

    \begin{tikzpicture}[scale = 0.7,>=stealth]
    \fill[color = lightgray] (5, 0) rectangle +(1, 0.5);
    \draw (0, 0) rectangle node {\scriptsize $\texttt{0}$} +(1, 0.5);
    \draw (1, 0) rectangle node {\scriptsize $\texttt{1}$} +(1, 0.5);
    \draw (2, 0) rectangle node {\scriptsize $\texttt{2}$} +(1, 0.5);
    \draw (3, 0) rectangle node {\scriptsize $\texttt{3}$} +(1, 0.5);
    \draw (4, 0) rectangle node {\scriptsize $\texttt{4}$} +(1, 0.5);
    \draw (5, 0) rectangle node {\scriptsize $\texttt{5}$} +(1, 0.5);
    \draw (6, 0) rectangle node {\scriptsize $\texttt{6}$} +(1, 0.5);
    \draw (7, 0) rectangle node {\scriptsize $\texttt{7}$} +(1, 0.5);
    \draw[->, thick] (2+0.5+4, 1-0.1) -- (2+0.5+4, 0.5);
    \node[scale = 0.8] at (2+0.5+0.2+4, 1+0.1) {$start: 6$};
    \draw[->, thick] (1+0.5+4, 1-0.1) -- (1+0.5+4, 0.5);
    \node[scale = 0.8] at (1+0.5-0.2+4, 1+0.1) {$end: 13$};
    \end{tikzpicture}
\end{center}
This state only represents the status of the tokens with 
indexes in $\{ 6, \cdots, 13 \}$, and the information of the
unused tokens with indexes $2$ and $3$ is lost (access requests originated from
these two tokens will be rejected).
Moreover, if an access request is made by a token with very
large index $i$ such that $i > end + n$, the bitmap will reset
all cells to zero and update the pointers according to~\autoref{alg:move}.

Therefore, the bitmap approach ensures that any one-time token can be used at
most once. It is possible that in certain situations some one-time tokens
become invalid before they are used, which is called a token miss.  For example,
if the smart contract has processed a token with the \texttt{index} 13, the
range of the bitmap is updated to \textit{start=6} and
\textit{end=13}. This implies that any (even unused) token with an
\texttt{index} smaller than 6 will be rejected (missed) by the contract.  
In this case, a holder of such an unused token would need to re-apply for a new
token from the TS again.  To avoid this situation, an owner should
allocate a large-enough bitmap in its smart contract.  There is a trade-off
between the size of the bitmap and the miss rate. The two factors that allow to
model the bitmap size are a) a token lifetime, and b) the (expected) maximum
number of transactions per second that the contract is going to process.  
Then the bitmap size required to not reject any unused and non-expired token is \textit{
token\_lifetime} $\times$ \textit{max\_tx\_per\_second} bits.  Fortunately, as we show
in~\autoref{subsect:costdiscussion}, even for the most popular Ethereum
contracts and realistic token lifetimes, the cost of the bitmap storage is low.

\begin{figure}[t]
\setlength{\belowcaptionskip}{-0.3cm}
    \centering
    \includegraphics[width=0.9\linewidth]{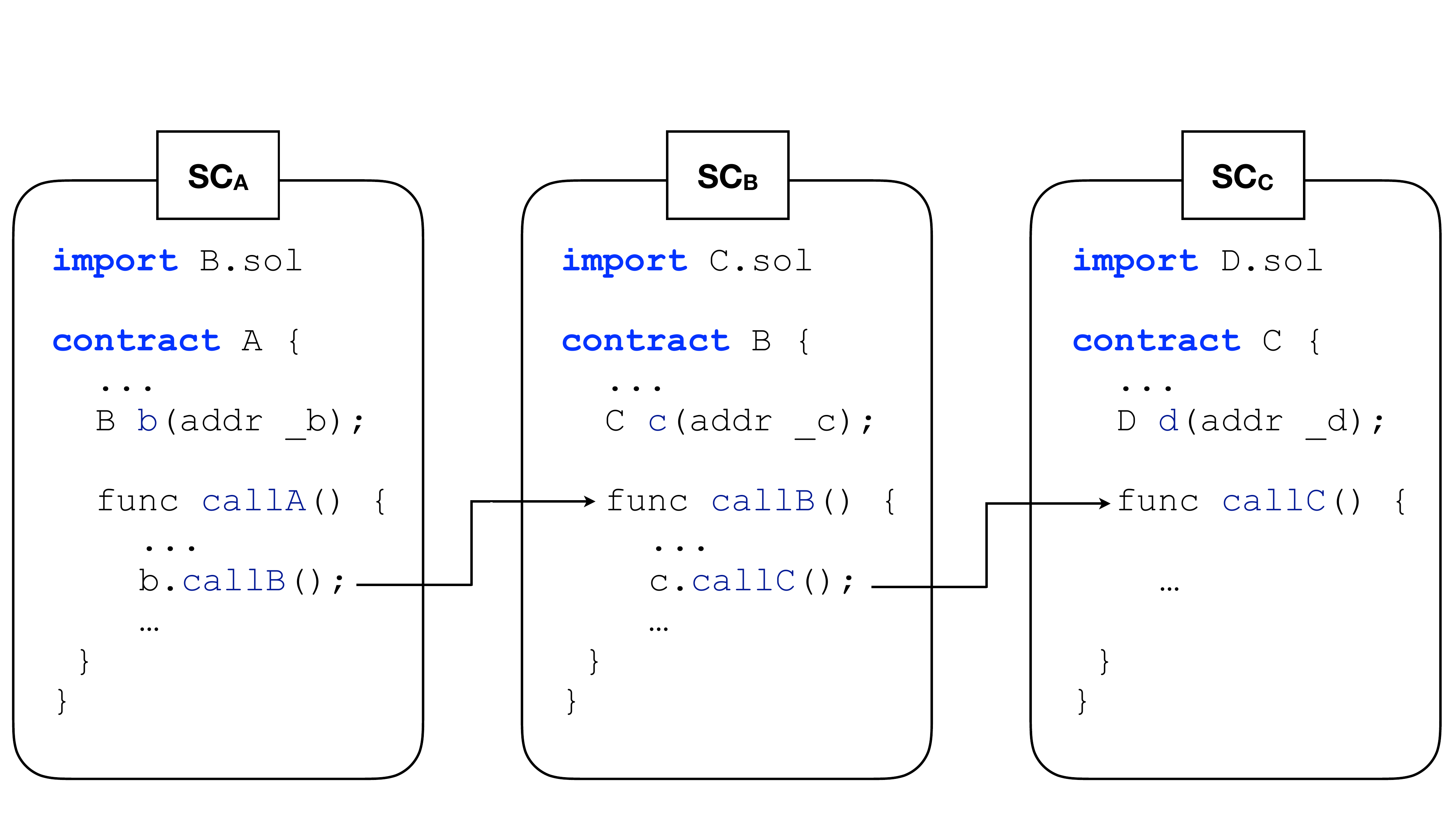}
    \caption{A call chain with the depth three.}
    \label{fig:multiple}
\end{figure}

\subsection{Tokens for Call Chains}
\label{subsect:multiple}
Starting from a transaction 
originated by an external owned account, an invoked contract method can 
call a method of another contract which in turn can call a method of a 
third contract, and this call chain (see~\autoref{subsec:txcall}) can go on.

When the smart contracts involved in the call chain are protected
with \smacs, the client initiating the call chain has to obtain 
proper tokens for all these smart contracts.
Let us consider a simple example shown in~\autoref{fig:multiple}.
Before a client triggers this call chain, she needs to
obtain three tokens (e.g., method tokens) from the TSes
corresponding to $SC_A$, $SC_B$, and $SC_C$ (these TSes can be operated
by different owners).

Assuming that the client successfully gets the three tokens
$tk_A$, $tk_B$, and $tk_C$, then she can embed an array of the three
tokens of the following form in the transaction:
{\setlength\abovedisplayskip{1pt}
\setlength\belowdisplayskip{1pt}
$$
SC_A:tk_A \parallel SC_B:tk_B \parallel SC_C:tk_C.
$$
When $SC_A$ receives the transaction,
it can extract the token ($tk_A$) associated with its address and verify 
validity according to~\autoref{alg:single}.
Subsequently, this array of tokens will be passed to 
$SC_B$ via a message call who parses out
$tk_B$ and verifies it. 
Finally, the array is passed while calling $SC_C$ which can perform the analogous operations as 
$SC_A$ and $SC_B$.

\subsection{Access Control Rules (\acr)}
Rules in \smacs define the set of token requests
which can successfully get the token from a TS when submitted.
For every token type, there is a set of rules associated
with it. A token request of a particular type will
be checked against the set of rules associated with that type.
In the following section, we present sample rules that can be implemented
in \smacs.

\paragraph{\textbf{Blacklist and Whitelist}}
Blacklist and whitelist are generic \acr supported in \smacs. 
In our context, the simplest form of a
whitelist is just a 
set of Ethereum addresses. 
Every address outside this list cannot
obtain a valid token from the TS, 
and therefore
it cannot access the \smacs-enabled smart contract. 
As depicted in~\autoref{fig:whitelist}, each token type 
has either a
blacklist or whitelist. 
\smacs does not mandate how these lists are created and for instance
an address whitelisted for super tokens
can be blacklisted for argument token. 
Moreover, the listed objects are not necessarily account addresses. For example, 
it is possible to blacklist dangerous argument values
for certain contract methods.
In \smacs, all these access lists can be dynamically updated by the
owner without any modification to the deployed smart contract.
\begin{figure}
\lstset{numbers=left,xleftmargin=2em}
\setlength{\belowcaptionskip}{-0.3cm}
    

    \begin{lstlisting}[language=json,firstnumber=1]
 {
   sender: {
     whitelist: ["0x366c...", "0xd488...",...],
   },
   method: {
     methodA: {
       blacklist: ["0xBa7F...", "0xb1D4...",...],
     },
     ...
   },
   argument: {
     argA: {
       whitelist: ["0x3540...", "0x9e9B...",...],
     },
     ...
   }
 }
 \end{lstlisting}
    \caption{An example of whitelists and blacklists.}
    \label{fig:whitelist}
\end{figure}

\paragraph{\textbf{Rules for Runtime Verification}}
Apart from these basic rules above,
the argument token type allows us to craft more advanced 
\acr that can enhance the \runtime security
of the \smacs-enabled smart contract.  
Imagining that a client tries to access a method of a smart contract
with a particular set of arguments. \smacs allows TSes to simulate the 
\runtime behavior of the smart contract in an isolated off-chain environment
and deny access if any abnormal behaviors triggered by the requested call are
observed. 
Then a TS implementing proper rules for argument tokens would be
able to protect even vulnerable already deployed smart contracts.
We show concrete instantiations of such rules in the next section.

%% file: sec/case.tex
Defensive logics with arbitrary complexity can be plugged into \smacs.
In particular, \smacs can be powerful when combined with 
any other \runtime verification tools preventing specific attack classes.
In this section, 
we show two
concrete examples where third-party tools are employed to implement
advanced rules enforcing certain \runtime security properties. 

\subsection{Enforcing Hydra Uniformity}
\label{subsect:hydra}
Hydra is a recent framework for smart contract bug bounty administration, which enables
\runtime detection and rewarding of critical bugs~\cite{Hydra}.  Basically, in
the Hydra framework, multiple independent program instances written in different
programming languages but with the same intended high-level logic run in parallel over
the blockchain. These program instances are called the heads (of the Hydra).

When a smart contract protected by Hydra executes, its intended logic 
proceeds normally only if the outputs of the heads are identical.
If the outputs diverge for different heads, it is likely that certain
erroneous state is triggered for some heads.
At this point, the execution of 
the smart contract aborts and the rewarding logic of Hydra takes control to pay out a 
bounty. Therefore, Hydra can detect bugs at \runtime at the cost of increased 
on-chain resource consumption by a factor round $N$ when $N$ heads are employed.

We integrated Hydra into \smacs by defining a dedicated rule
for argument tokens. This rule dictates that an argument token
is issued only when the outputs of all heads are identical
when called with the payload specified in the token request.  
In contrast to Hydra, heads in \smacs are run by a TS on its local testnet. 
Hydra acts as a simulator in \smacs, does not consume on-chain resources,
and therefore it is possible to implement more heads in our case without
introducing additional on-chain cost.
In summary, this rule enforces Hydra uniformity, where transactions 
leading to different head outputs are unable to get a token at the first
place.
We show the efficiency of Hydra-supported \smacs in~\autoref{subsect:performance}.

\subsection{Blocking Re-entrancy Attacks}
\label{subsect:reentrancy}
In this case, we show how to protect a smart contract from 
the so-called \textit{re-entrancy attack}, the essence of 
the real-world TheDAO attack~\cite{grossman2017online}, 
leading to a loss of over \$50 million
worth of Ether at the time the attack occurred. 

Let us consider the vulnerable smart contract \texttt{Bank}, a simplified version of TheDAO~\cite{daodisco}, as shown in~\autoref{fig:bank}.
Anyone can deposit ether into \texttt{Bank}, the 
amount of ether is recorded in the mapping \texttt{balance}.
The ether deposited can be withdrawn by calling \texttt{withdraw()},
which sends the ether to the \texttt{msg.sender} address.
This transfer implicitly triggers a \textit{fallback method} (an
anonymous method that does not take any arguments) of the
receiver. This default behavior can have security consequences as the execution
flow can be controlled by a remote fallback method. 
\begin{figure}[t!]
\setlength{\belowcaptionskip}{-0.8cm}
\lstset{numbers=left,xleftmargin=2em}

	\begin{lstlisting}[language=Solidity]
contract Bank{
	mapping(address=>uint) balance;
	function addBalance() public{
		balance[msg.sender] += msg.value;
	}
	function withdraw() public{
    uint amount = balance[msg.sender];
		if (msg.sender.call.value(amount)() == false) {throw;}
		balance[msg.sender] = 0;
	}
}

contract Attacker{
	bool isAttack; address bank;
	function Attacker(addr _bank, bool _isAttack){
		bank = _bank; isAttack = _isAttack;
	}
	function() payable{
		if(isAttack == true){
			isAttack = false;
			if(bank.withdraw()) {throw;}
		}
	}
	function deposit(){
		bank.call.value(2).addBalance();
	}
	function withdraw(){
		bank.withdraw():
	}
}
	\end{lstlisting}
    \caption{The \texttt{Bank} contract with a re-entrancy vulnerability and the
    \texttt{Attacker} contract exploiting it.}
	\label{fig:bank}
	\label{fig:attacker}
\end{figure}
The re-entrancy attack can be lunched by an attacker using the 
smart contract shown in~\autoref{fig:bank}. She first calls 
the \texttt{deposit()} method to deposit two ethers into the 
target smart contract \texttt{Bank}.
Now she is ready to attack the target by calling the \texttt{withdraw()}
method of \texttt{Attacker}. Subsequently, 
\texttt{Attacker.withdraw()} calls \texttt{Bank.withdraw()} which
then triggers a recursive \texttt{Bank.withdraw()} call
via \texttt{Attacker}'s fallback method,
and the line 11 of the \texttt{Bank} smart contract is never reached.
The above attack strategy effectively moves all ether from \texttt{Bank}
to the account controlled by the attacker.

To prevent \texttt{Bank} from being exploited, we use \smacs with a rule
employing ECFChecker~\cite{ecf} -- a developed tool for
detection of \textit{effectively callback free} objects~\cite{grossman2017online}.  
To integrate that, the TS deploys an ECFChecker-supported
implementation running an off-chain testnet with the \texttt{Bank} contract
deployed. For every token request, the TS calls a requested method with the passed
arguments and observes the output of ECFChecker. 
The TS issues the tokens only if ECFChecker does not report any security issue.
We emphasize that the described integration gives the contract owner
ECFChecker security
benefits without requiring Ethereum participants to update their 
configurations to support ECFChecker.
In~\autoref{subsect:performance} we show the efficiency of this setup.

%% file: sec/evaluation.tex
To evaluate our design, we fully implement
the \smacs framework.
\smacs-enabled smart contracts are developed
by \texttt{Solidity v0.4.24}
and deployed on a testnet. 
The TS is implemented as a web server running
\texttt{Node.js v10.2.1} bundled 
with the \texttt{node-localStorage} package 
for storing rules and signature key-pairs. 
We implement client and owner with
\texttt{web3.js}~\cite{web3}.
This
software interacts with deployed \smacs-enabled smart contracts and
TSes.
We use the Ethereum's ECDSA signature scheme as the default one, as Ethereum
provides a native and optimized support for it.

\subsection{Gas Cost}
\label{subsect:costdiscussion}
In the \smacs framework, clients send transactions with proper tokens which
are verified by smart contracts. Therefore, the main cost is introduced with
respect to the computation and storage whose utilization 
is charged by the Ethereum network. We perform a series of experiments to measure the cost 
introduced by \smacs in terms of gas consumption.

We conduct experiments for different types of tokens and record their gas cost, 
together with the cost converted to US dollars
in~\autoref{table:atomicgascost}.
The conversion was according
to the gas price from~\cite{gasstation} at the time of writing
the paper.
From the table, we can see that the 
dominating operation is the signature verification.
The cost also increases in
arguments tokens as they require more
processing (\texttt{argName} and \texttt{argValue} have to be processed).
However, the overall cost of a token verification is around \$0.04 for super and
method tokens and around \$0.1 for argument tokens.
As shown in the table, for tokens with the one-time property the verification
gas consumption is similar, despite 
additional operations are required by the bitmap.

\begin{table}[!t]
\setlength{\belowcaptionskip}{-0.1cm}
\setlength{\abovecaptionskip}{0.1cm}
  \caption{Single token processing gas cost.}
  \label{table:atomicgascost}\centering
\begin{tabular}{lrrr}
\toprule
\multirow{2}*{Cost} & \multicolumn{3}{c}{Token type (without the one-time property)}\\ \cmidrule{2-4}
          & \texttt{Super}  &\texttt{Method}  &\texttt{Argument} \\ \midrule
Verify    &108282 (65\%)  &115108 (67\%)  &330889 (85\%)  \\ 
Misc      &57675 (35\%)   &57675 (33\%)   &57678 (15\%)   \\\bottomrule
Total     &165957   &172783    &388567    \\
USD       &0.041            &0.042            &0.094            \\ \bottomrule
\bottomrule
\multirow{2}*{Cost} & \multicolumn{3}{c}{Token type (with the one-time property)}\\ \cmidrule{2-4}
          & \texttt{Super}   &\texttt{Method}  &\texttt{Argument}\\ \midrule
Verify    &108531 (56\%)   &115651 (58\%)  &330914 (79\%) \\ 
Misc      &57426  (30\%)   &56994 (28\%)   &57331  (14\%) \\
Bitmap    &27471 (14\%)    &27839 (14\%)   &28003 (7\%)  \\ \bottomrule
Total     &193428     &200484    &416248  \\
USD       &0.047             &0.048            &0.101   \\ \bottomrule
\end{tabular}
\end{table}
\begin{table}[t!]
\setlength{\belowcaptionskip}{-0.1cm}
\setlength{\abovecaptionskip}{0.05cm}
  \caption{Gas cost for multiple one-time argument tokens.}
  \label{table:multiplegascost}\centering
\begin{tabular}{L{0.36cm}R{1.64cm}R{1.64cm}R{1.64cm}R{1.78cm}}
\toprule
\multirow{2}*{Cost} & \multicolumn{4}{c}{Number of Token}\\ \cmidrule{2-5}
          & \texttt{1}      &\texttt{2}       &\texttt{3}      &\texttt{4} \\ \midrule
Verify    &330914 (79\%)    &662952 (79\%)    &994552 (78\%)   &1326506 (78\%)  \\ 
Misc      &57331 (14\%)     &102991 (12\%)    &150463 (12\%)   &203499 (12\%)  \\
Bitmap    &28003 (7\%)      &56746 (7\%)      &84612 (7\%)     &112034 (7\%)    \\
Parse     &--               &16986 (2\%)      &34182 (3\%)     &57872 (3\%)    \\\bottomrule
Total     &416248    &839675    &1263809  &1699911  \\
USD       &0.101            &0.204            &0.307           &0.412   \\ \bottomrule
\end{tabular}
      \vspace{-0.2cm}
\end{table}

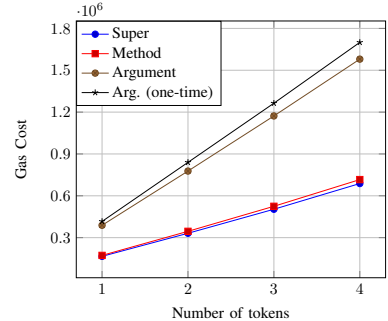
\begin{figure}[t!]
\setlength{\abovecaptionskip}{0.1cm}
\setlength{\belowcaptionskip}{-0.3cm}
  \centering
\begin{tikzpicture}[scale = 0.6,>=stealth]
\begin{axis}[
    legend cell align={left},
    xtick={0,1,...,4},
    ytick distance=300000,
    grid=major,
    yticklabel style={
            /pgf/number format/fixed,
            /pgf/number format/precision=2,
            /pgf/number format/fixed
    },
    legend style={
    at={(0,0)},
    anchor=north west,at={(axis description cs:0,1.0)}},
    xlabel={Number of tokens},
    ylabel={Gas Cost}
]
\addplot coordinates {
    (1,165957) (2,331365) (3,503520) (4,687941)
};
\addplot coordinates{
    (1,172783) (2,345345) (3,524782) (4,716207)
};
\addplot coordinates{
    (1,388567) (2,777309) (3,1172391) (4,1579761)
};
\addplot coordinates{
    (1,416248) (2,839675) (3,1263809) (4,1699911)
};

    \legend{Super,Method,Argument,Arg. (one-time)}
\end{axis}
\end{tikzpicture}
  \caption{Aggregated gas cost for verifying multiple tokens.}
  \label{fig:all}
\end{figure}

As discussed in~\autoref{subsect:multiple}, 
\smacs supports transactions that invoke a call chain of contracts. 
In this case, 
the token verification cost varies according to the depth of the chain, 
and additional cost is induced since a contract has to parse
the passed token array before verification.
We conduct analogical experiments as in the previous case
and the results are shown in~\autoref{table:multiplegascost}
and~\autoref{fig:all}.
(Note that the table presents the results for the argument token type whose
verification is around two times more gas consuming than other types.)
As presented, the verification cost increases linearly with the call chain
length.

Implementing the one-time property requires to store a bitmap by smart
contracts.
The size 
of this storage depends on the token
lifetime and the expected transaction frequency, however, 
this cost is one-time, paid upon the contract creation.
To give insights on the cost we take the
ten most popular smart contracts based on the number of transactions by Jan,
2019~\cite{popularcontracts} and analyze their transactions
distribution. We found that on average the transaction peak is \textit{35 tx/s}
which is close to the Ethereum's maximum
throughput~\cite{maxthroughput}.
Setting the lifetime of one-time tokens to one hour and assuming conservatively 
that all transactions use one-time tokens,~\autoref{table:storagecost} shows
the required storage and its cost.  We can see that to handle even \textit{35 one-time
tokens/s} a smart contract has to be initialized with storage
costing only one-time fee of \$2.14.  This cost is linear in transaction
frequency and token lifetime.

\subsection{\ts Performance}
\label{subsect:performance}
\begin{figure}[t!]
\setlength{\belowcaptionskip}{-0.3cm}
\setlength{\abovecaptionskip}{0.1cm}
  \centering
\begin{tikzpicture}[scale = 0.6,>=stealth]
\begin{axis}[
    legend cell align={left},
    ymin=50,
    ymax=350,
    grid=major,
    xmode=log,
    ytick distance=100,
    yticklabel style={
            /pgf/number format/fixed,
            /pgf/number format/precision=2,
            /pgf/number format/fixed
    },
    legend style={
    at={(0,0)},
    anchor=north west,at={(axis description cs:0,1.0)}},
    scaled y ticks=false,
    xlabel={Number of requests},
    ylabel={Requests processed per second}
]
\addplot coordinates {
    (1,123) (1e1,143) (1e2,165) (1e3, 221) (1e4,248)  (1e5,252)
};
\addplot coordinates{
    (1,110) (1e1,123) (1e2,145) (1e3,197) (1e4,230) (1e5,233)
};
\addplot coordinates{
    (1,107) (1e1,113) (1e2,139) (1e3,192) (1e4,196) (1e5,202)
};
\addplot coordinates{
    (1,84) (1e1,96) (1e2,112) (1e3,126) (1e4,143) (1e5,150)
};

    \legend{Super,Method,Argument,Arg. (one-time)}
\end{axis}
\end{tikzpicture}
  \caption{Throughput of the TS.}
  \label{fig:throughput}
\end{figure}
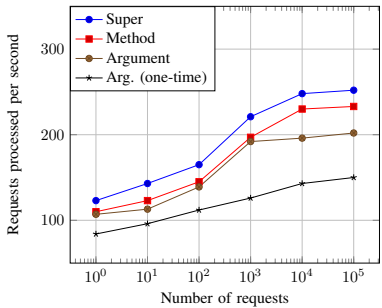

\paragraph{\textbf{TS Throughput}}
We evaluate the TS throughput running a TS instance on a system with macOS Sierra 10.12.6,
Intel Core i5 CPU (2.7 GHz), and 8GB RAM.
For each token type, we send $10^i$ ($0 \leq i \leq 5$) token requests to the TS,
record the total time needed by the TS
, and compute 
the average time required per token request.
The rules used are composed of blacklists and whitelists as presented
in~\autoref{fig:whitelist}.
The obtained results are summarized in~\autoref{fig:throughput}.

\begin{table}[t!]
    \caption{Storage cost for the bitmap (this cost is one-time).}
  \label{table:storagecost}\centering
\begin{tabular}{lrrrr}
\toprule
\multirow{2}*{Cost} & \multicolumn{3}{c}{Transaction frequency (tx/s)}\\ \cmidrule{2-4}
          & \texttt{35}   &\texttt{3.5} &\texttt{0.35} \\ \midrule
Storage     &15.38 KB  &1.54 KB  &0.154 KB          \\
Deployment    &8849037  &886054  &88605         \\
USD       &2.140          &0.214          &0.021   \\ \bottomrule
\end{tabular}
      \vspace{-0.2cm}
\end{table}

From~\autoref{fig:throughput} we can see the number of
token requests handled per second raises 
when the requests are processed in batches.
The throughput becomes stable when the number of requests is greater than $10^5$,
with the 
time cost about 5ms for most token types.
The single TS instance can easily handle
all transactions processed by the current Ethereum main network even in peak times.
We found the ever highest transactions peak in Ethereum
for one of the most popular 
smart contracts  -- CryptoKitties~\cite{kittyaddress}
when it received about 48 transactions per second (on 05-Dec-17 00:43:03 UTC~\cite{blockspur,etherscan}).

\paragraph{\textbf{Integration with Runtime Tools}}
In~\autoref{sec:case}
we integrate \smacs
with two \runtime verification tools, i.e., Hydra and ECFChecker.  
In both cases, TS to verify incoming requests requires \textit{local} Ethereum
testnets (no code changes at the deployed
contracts or Ethereum software used by network nodes are required, however).
To improve throughput of the tools we
configure the testnets (i.e., geth~\cite{geth}) to minimize the latency between
submitting and executing transactions.
For Hydra, we implement a simple contract in three different programming
languages and deploy it on a Hydra-supported testnet.
For ECFChecker, we deploy the vulnerable contract presented in~\autoref{sec:case}.
We send 100 transactions each and measure the
average time needed by a tool to process a transaction.  
In our setting, \smacs with Hydra needs 120ms to process a request,
while ECFChecker-supported \smacs can process a request in
only 10ms.
Thus, with these tools \smacs can handle around 8 and 100 token requests
per second.


%% file: sec/disc.tex
\subsection{Security}
Our first claim is that an adversary cannot bypass the access control in \smacs.
All contract's public interfaces require the token verification process (see~\autoref{subsec:token_issue_verify}).  This process ensures that a token is
authentic (i.e., signed by the TS), non-expired, and matches the calling
transaction.  The only way of obtaining such a token is to request the TS which
would verify the request against its access rules.  All valid tokens are
signed by the TS, therefore an adversary without passing the TS validation or
without the TS' private key cannot get a valid token for its transactions. Moreover, 
one-time tokens could be issued only the clients satisfied the rules predefined in whitelist,
guaranteeing one-time access even if an adversary controls multiple addresses.

\paragraph{\textbf{Substitution Attack}}
An attacker can intercept a transaction from a legitimate client,
extract the token from it, and then construct a transaction with
the intercepted token. This transaction will be rejected by the 
contract-side verification, since the signature of the TS creates a cryptographic 
binding of the token and the context in which the token can be applied.
Any tiny change of the context (e.g., address, argument, etc.) will be 
caught by the signature verification process.

\paragraph{\textbf{Replay Attack}}
An attacker can capture the transaction originated from a legitimate client, and
replay it in the Ethereum network.  This attack is against Ethereum itself and
cannot succeed since the built-in countermeasure of Ethereum against replay
attack will reject the replayed transaction. The \textit{nonce} value included in
the transaction ensures that every single transaction is unique. If a
transaction has been accepted by Ethereum, it will not be processed again.
Moreover, the client's address is protected by a token's signature, thus, the
attacker cannot extract and reuse others' tokens.
\smacs implements in-contract replay protection for one-time tokens by assigning
each such a token with a unique index (set by the TS) and recording every use of
a one-time token in the stored bitmap.  A client can try to replay a one-time
token by creating a new transaction with the used token embedded.  Such a
transaction will not trigger an actual execution of the targeted method, as
the token verification procedure (see~\autoref{subsect:onetime}) will check
whether the token index was already used, and in that case deny access.

\paragraph{\textbf{51\% Attack}}
In the 51\%-attack an adversary possesses more than 50\% of the total voting
power of the blockchain network, what allows her to rewrite the blockchain
history.  This kind of attacks is devastating as they allow to double-spend,
however, in our context even such a strong adversary cannot bypass the \smacs
access control.  The adversary can disorder or even remove transactions at will,
compromising the availability of smart contracts, but she cannot obtain a
valid token for a non-compliant transaction.

\paragraph{\textbf{Privacy}}
\smacs moves access rules to TSes which are off-chain services. Therefore
deployed rules, verification tools, and their configurations are kept
private and are not
revealed even to clients.
As blockchain transactions are publicly visible, an adversary can learn
successful access control cases and try to predict the applied rules, however
it is still a black-box analysis (in contrast to any in-contract access
control).

\subsection{Deployment}
\paragraph{\textbf{Availability}}
Requiring a TS to keep verifying and signing tokens introduces a single point of failure, as
with the failed TS clients would not be able to interact with the contract.
Fortunately, TSes in \smacs are easy to scale and replicate. 
For issuing tokens (without the one-time property) providing availability is as
easy as providing redundant TSes that do not require any coordination (except a 
load-balancer/failover system). 
If a TS service is offering one-time tokens, then its replicas have to coordinate on
the current counter value (see~\autoref{subsect:onetime}). That can be
efficiently realized via
a replicated counter primitive usually implement upon a standard consensus
algorithm~\cite{burrows2006chubby,ongaro2014search}.

\paragraph{\textbf{Service Discovery}}
We implicitly assumed that clients know how to reach the TS corresponding to a
\smacs-enabled smart contract.
In practice, clients have to learn an URL address of the service. We propose to
implement this discovery process by adding the service address as a smart
contract instance metadata (similarly as contract's name or the compiler version it was
created with).

%% file: sec/related.tex
In practice, 
the community has
developed some design patterns and even third-party libraries to
facilitate the application of access control 
over smart contracts~\cite{OpenZeppelin}.
However, this paradigm puts the burden of all access control logic
and its management on the smart contract itself. 
Due to the high cost of on-chain resources,
only simple and inflexible \acr can be developed using this 
approach (e.g., a blacklist or whitelist with small size, a role-based
\acr supporting a very limited number of roles, etc.). 
In summary,
although smart contracts access control is an obvious need
and an important aspect of smart contract security which 
has been extensively investigated over the last years, we are
not aware of any framework similar to \smacs, which could 
implement complex \acr supporting 
\runtime security verification at a very low cost.
The most relevant research to \smacs is 
the investigation and development of methodologies and tools for detecting 
vulnerabilities of smart contracts, which can be divided into
two general categories: \textit{static} and 
\textit{\runtime} security analysis.

\paragraph{\textbf{Static Security Analysis}}
These methods or tools mainly based on formal verification 
and symbolic execution. 
Oyente~\cite{luu2016making} and Manticore~\cite{Manticore} are 
symbolic execution tools for finding potential security bugs.
Mythril~\cite{Mythril} uses concolic analysis, taint analysis, and control 
flow checking to detect multiple smart contract security vulnerabilities.
Securify~\cite{tsankov2018securify} extracts semantic facts 
by performing advanced static analysis to 
prove the presence or absence of certain security vulnerabilities. 
Zeus~\cite{kalra2018zeus} employs model checking to verify the 
correctness of smart contracts.
MAIAN~\cite{MAIAN},
processes the bytecode of smart contracts and tries to build a trace of 
transactions to find and confirm bugs based on inter-procedural symbolic analysis.
The list of tools are difficult to enumerate and 
new relevant tools are constantly emerging~\cite{grishchenko2018foundations,ducasse2019smartanvil,solgraph,verx}. 
Most of these tools 
are meant to provide pre-deployment security verification.
Thus they can only identify bugs (rather than protect from them) for already deployed smart contracts.

Another drawback is that it cannot fully cover
all \runtime behaviors and therefore is susceptible to missing 
novel \runtime attack patterns. In fact, this has been demonstrated
in~\cite{rodler2018sereum}, where 
new re-entrancy attack vectors are crafted
which bypass the security check of existing static analysis tools~\cite{kalra2018zeus,luu2016making}.  

We see this class of tools as orthogonal to \smacs, however, we believe that
in some cases they could be used in combination providing security benefits.
For example, the owner of a \smacs-enabled smart contract can
scan the deployed contract regularly with such tools (e.g., perform
a vulnerability scan whenever the security analysis tools get updated).
Once a vulnerability is detected, she can blacklist transactions with specific patterns that can
potentially trigger the vulnerability.

\paragraph{\textbf{Runtime Security Analysis}}
In contrast to static security analysis, 
tools~\cite{grossman2017online,breidenbach2018enter,rodler2018sereum} 
performing \runtime monitoring
has the potential to prevent deployed smart contracts from being exploited.
Hydra~\cite{breidenbach2018enter} enables post-deployment
security through $N$-of-$N$-version programming, a variant of classical
$N$-version programming that runs multiple
independent program instances to detect \runtime security issues.
ECFChecker~\cite{grossman2017online} is a runtime
detection tool dedicated to finding \textit{effectively callback free} objects.
This tool can be used for finding Ethereum re-entrancy attacks.
More detailed overview of Hydra and ECFChecker can be found in~\autoref{sec:case}.

Another interesting example is the Sereum~\cite{rodler2018sereum} architecture,
a hardened EVM which is able to protect deployed
contracts against re-entrancy attacks in a backward compatible way
by leveraging taint tracking to monitor \runtime behaviors of smart contracts.
Sereum can also be integrated into the \smacs framework easily by 
using dedicated \acr.

The main drawback of these tools is their requirement of changing and upgrading
the runtime environment. We emphasize that in the replicate state machine model
followed by blockchain platforms, this implies that a majority of nodes would
need to update their EVMs to support such a tool. 
\smacs enables contract owners to benefit from these tools without this
requirement.  Moreover, as we presented, these tools can be easily and
seamlessly integrated with \smacs.
Another preferable feature offered
by combining \runtime security analysis tools with \smacs is that 
a vulnerable smart contract may still operate normally, since only
innocent transactions pass through and suspicious transactions identified
by the tools are rejected at \runtime.

%% file: sec/conclusions.tex
We presented \smacs,
to the best of our knowledge, the first framework that
achieves efficient, flexible, and fine-grained access control of
smart contracts with low cost by combining lightweight on-chain 
verifications and off-chain access control management infrastructures.
%
Apart from enabling malicious addresses prevention and abnormal 
\runtime behaviors resistance for smart contracts,
\smacs offers several preferable features.
Firstly, when combined with \runtime verification tools, 
a \smacs-enabled smart contract can deny suspicious access attempts 
on the fly while 
keeping operating for innocent transactions.
Secondly, the architecture of \smacs allows rules for enhancing 
{\textit{post-deployment} security to be designed based on which it is possible
to prevent vulnerabilities discovered \textit{after deployment} from being exploited.
Therefore, it is meaningful to test \smacs-enabled contracts with new
verification tools
regularly and adjust the rules accordingly. 
Finally, due to the extensibility of the framework, 
we could expect more security-related tools that can be applied 
in \smacs to emerge in the future. 

An interesting research direction is to investigate trusted execution
environments (TEEs, e.g., Intel SGX) in the context of \smacs to fully decentralize it. 
For instance, a TS implemented
within a TEE enclave could decentralize the entire system: an owner would
just publish its \acr which would be validated by the enclave code
running locally on a client (without contacting any central service).
We leave a detailed design of such a system as future work.